\documentclass[12pt,a4]{article}
\usepackage[ansinew]{inputenc}
\usepackage{graphicx}
\usepackage{color}
\usepackage[colorlinks]{hyperref}
\usepackage[active,new,noold,marker]{xrcs}
\usepackage{graphics}
\usepackage{latexsym}
\usepackage{showidx}
\usepackage{amssymb}
\usepackage{amsfonts}
\usepackage{amsmath,amsthm}

\font\msbm=msbm10 at 12pt
\newcommand{\R}{\mbox{\msbm R}}

\newcommand{\Z}{\mbox{\msbm Z}}

\newcommand{\F}{\mbox{\msbm F}}

\title{Construction of LDGM lattices}
\author{
Hassan Mehri\footnote{Corresponding Author.{\it E-mail
 Address:}hassanmehri\textbf{.}math@gmail.com} and Mohammad Reza Sadeghi
\\
[5mm]
{\it \small Faculty of Mathematics and Computer Science,}\\
{\it \small Amirkabir University of Technology,}\\
{\it \small No.424, Hafez Avenue, Tehran 15875-4413, Iran,} }
\begin{document}
\maketitle
%%%%%%%%%%%%%%%%%%%%%%%%%%%%%%%%%%%%%%%%%%%%%%%%%%%%%%%%%%%%%%
\vspace*{0cm}
\begin{abstract}
Low density generator matrix ($LDGM$) codes have an acceptable
performance under iterative decoding algorithms. This idea is
used to construct a class of lattices with relatively good
performance and low encoding and decoding complexity. To
construct such lattices, Construction \textit{\textbf{D}} is
applied to a set of generator vectors of a class of $LDGM$ codes.
Bounds on the minimum distance and the coding gain of the
corresponding lattices and a corollary for the cross sections and
projections of these lattices are provided. The progressive edge
growth (PEG) algorithm is used to construct a class of binary
codes to generate the corresponding lattice. Simulation results
confirm the acceptable performance of these class of lattices.
\end{abstract}
%%%%%%%%%%%%%%%%%%%%%%%%%%%%%%%%%%%%
\smallskip
\noindent {\bf Index Term:} Lattice, PEG algorithm, $LDGM$ codes.
%%%%%%%%%%%%%%%%%%%%%%%%%%%%%%%%%%%%
\section{INTRODUCTION}
\label{sec:Intro} The lattice version of the Gaussian channel
coding problem for a given value of signal to noise ratio (SNR)
is to find the n-dimensional lattice for which the error
probability is minimized \cite{Conway}. It is shown that,
lattices can achieve the capacity of additive white Gaussian
noise (AWGN) channel \cite{forneyspherebound,forneyconf}. This
fact motivates the search for lattices with large coding gains.
On the other hand in larger dimensions the encoding and decoding
complexity also increase. There are several methods to construct
lattice from linear codes \cite{Conway}. Among them, Construction
$\textit{{\textbf{D}}}$ and Construction
$\textit{{\textbf{D}}}^{'}$ can produce high coding gain lattices
by using a collection of linear codes. The idea of Low density
generator matrix codes were first provided by Garcia and Zhong
\cite{Garcia_2003}. In addition to low encoding and decoding
complexity, these linear codes have relatively good performance.
As a result, constructing lattices based on these codes can be a
promising tool. Therefore we will propose a class of lattices
with almost high coding gain and low encoding and decoding
complexity. The paper begins in the next section with a brief
discussion about lattice. Section three introduces the
Construction \emph{$\textbf{D}$} lattices. Systematic low density
generator matrix lattices discussed in the forth section. The
final section is dedicated to the paper's conclusions.
%%%%%%%%%%%%%%%%%%%%%%%%%%%%%%%%%%%%%%%%
\section{BACKGROUND}
\label{Sec:Background} Low density generator matrix ($LDGM$) codes
are linear codes which have sparse generator matrix
\cite{Garcia_2003}. Let ${\R}^{m}$ be the $m$-dimensional real
vector space with the standard product $\langle.,.\rangle$ and
Euclidean norm
$\parallel\textbf{x}\parallel=\langle\textbf{x},\textbf{x}\rangle^{1/2}$.
%A lattice $\Lambda$ is a discrete additive subgroup of ${\R}^{m}$.
%An $n$-dimensional lattice is generated by the integer
%combinations of a set of $n$-linearly independent vectors
An $n$ dimensional lattice in ${\R}^{m}$ is defined as the set of
all linear combinations of a given basis of $n$ linearly
independent vectors in ${\R}^{m}$ with integer coefficients
\cite{Conway}. Any subgroup of a lattice $\Lambda$ is called
sublattice of $\Lambda$ and a lattice is called
\textit{orthogonal} if it has a basis with mutually orthogonal
vectors. The set $\Lambda^{*}$ of all vectors in the real span of
$\Lambda$ (\textit{span}($\Lambda$)), whose the standard inner
product with all elements of $\Lambda$ has an integer value, is an
$n$-dimensional lattice called the \textit{dual} of $\Lambda$.
Lattices constructed by Construction $\textit{\textbf{D}}$ have a
square generator matrix thus if $\textbf{B}$ is a generator
matrix for $\Lambda$, then $\textbf{B}^{*}=\textbf{B}^{-1}$ is a
generator matrix for $\Lambda^{*}$ (parity-check matrix of
$\Lambda$). Every lattice point is therefore of the form
$\textbf{v}=\textbf{B}\textbf{x}$ where $\textbf{x}$ is an
$n$-dimensional vector of integers. The Voronoi cell of a lattice
point is defined as the set of all points in ${\R}^{m}$ that are
closer to this lattice point than to other lattice point. The
Voronoi cells of all lattice points are congruent and for
Lattices constructed by Construction $\textit{\textbf{D}}$ the
volume of the Voronoi cell is equal to volume of $\Lambda$
\cite{Conway}. The coding gain of lattice $\Lambda$ is defined by
\begin{equation}\label{eq:1}
\gamma(\Lambda):=
\frac{\textit{d}^{2}_{min}(\Lambda)}{{det(\Lambda)}^{2/n}},
\end{equation}
where $\textit{d}_{min}(\Lambda)$ and $det(\Lambda)$ refer to
minimum distance and volume of $\Lambda$, respectively
\cite{Conway}. Assume an $n$-dimensional lattice $\Lambda$ with an
$n$-dimensional orthogonal sublattice $\Lambda^{'}$ which has a
set of basis vectors along the orthogonal subspace
$S=\{W_{i}\}^{n}_{i=1}$. By the definition of the projection onto
the vector space $W_{i}$ as $P_{W_{i}}$ and the cross section
$\Lambda_{W_{i}}$ as $\Lambda_{W_{i}}=\Lambda \cap W_{i}$. Now,
the \emph{label groups} $G_{i}$ is defined as
$G_{i}=P_{W_{i}}/\Lambda_{W_{i}}$, which is used to label the
cosets of $\Lambda^{'}$ in $\Lambda$. Let $|G_{i}|=g_{i}$ and
$\textbf{v}_{i}$ be the generator vector of $\Lambda_{W_{i}}$,
\textit{i.e.}, $\Lambda_{W_{i}}=\Z \textbf{v}_{i}$. Each element
of $G_{i}$ can be rewritten in the form of
$\Lambda_{W_{i}}+j\det(P_{W_{i}})\textbf{v}_{i}/|\textbf{v}_{i}|(j=1,\ldots,g_{i}-1)$.
Then the map
\begin{equation}\label{eq:02}
\Lambda_{W_{i}}+j\det(P_{W_{i}})\frac{\textbf{v}_{i}}{|\textbf{v}_{i}|}\longrightarrow
j
\end{equation}
is an isomorphism between $G_{i}$ and ${\Z}_{g_{i}}$, thus every
element of the label group $G_{i}$ can be written as
$(\Z+a_{j})\textbf{v}_{i}$, where
$a_{j}=j~det(P_{W_{i}})/det(\Lambda_{W_{i}})$
\cite{banihashemi2001}.
%%%%%%%%%%%%%%%%%%%%%%%%%%%%%%%%%%%%%%%%%%%%%%%%%%%%%%%%%%%%%%%%%%%%%%%%%%%%%%%%%%%%%%%%%%%%%%%
\section{CONSTRUCTION \emph{$\textbf{D}$} LATTICES}
\label{Sec:CONSTRUCTIOIN}
%%%%%%%%%%%%%%%%%%%%%%%%%%%%%%%%%%%%%%%%%%%%%%%%%%%%%
Between other constructions of lattices from linear codes,
Construction $\textit{\textbf{D}}$ seems to be one of the best
choices for constructing lattices from $LDGM$ codes
\cite{sadeghi_refrence_5}. This construction can produce lattices with high coding gains and it deal with generator sets of codes.\\
\\
Let $\alpha=1$ or $\alpha=2$ and\ $C_{0} \supseteq C_{1}\supseteq
\ldots C_{a}$ be a family of binary linear codes, where the code
$C_{l}$ has parameters $[n,k_{l},d^{(l)}_{min}]$ with
$d^{(l)}_{min}\geq4^{l}/\alpha$, for $l=1,\ldots,a$ and $C_{0}$
is the trivial code $\F^{n}_{2}$. Choose a basis
$\{\textbf{c}_{1},\ldots,\textbf{c}_{n}\}$ for $\F^{n}_{2}$ such
that
$C_{l}=\langle\textbf{c}_{1},\ldots,\textbf{c}_{k_{l}}\rangle$.
For any element $\textbf{x}=(x_{1},\ldots,x_{n})\in \F^{n}_{2}$
consider
\[
\frac{\textbf{x}}{2^{l-1}}=(\frac{x_{1}}{2^{l-1}},\ldots,\frac{x_{n}}{2^{l-1}})
\]
as a vector in $\R^{n}$. Then $\Lambda \subseteq \R^{n}$ consists
of all vectors of the form
\begin{equation}\label{eq:03}
\textbf{z}+\sum_{l=1}^{a}\sum_{j=1}^{k_{l}}\frac{\alpha^{(l)}_{j}}{2^{l-1}}\textbf{c}_{j},
\end{equation}
where $\textbf{z}\in(2\Z)^{n}$ and $\alpha^{(l)}_{j}=0,1$.\\
\\
\emph{Theorem 3.1:} The set $\Lambda$ is a lattice, with minimum
distance at least 4/$\alpha$, determinant
\begin{equation}\label{eq:04}
det(\Lambda)=2^{n-\sum^{a}_{l=1}k_{l}},
\end{equation}
and coding gain of $\Lambda$ is
\begin{equation}\label{eq:05}
\gamma(\Lambda)\geq\alpha^{-1}4^{\sum^{a}_{l=1}\frac{k_{l}}{n}}.
\end{equation}
An integral basis for $\Lambda$ is given by the vectors
\[
\frac{1}{2^{l-1}}\textbf{c}_{j}\quad for\quad l=1,\ldots,a\quad
and\quad j=k_{l+1}+1,\ldots,k_{l},
\]
plus $n-k_{1}$ vectors of the form $(0,\ldots,0,2,0,\ldots,0)$.\\
\\
The proof is given in the appendix.\\
\\
\textit{Corollary 3.1:} Let $\textbf{B}$ be the generator matrix
of the lattice constructed using Construction
$\textit{\textbf{D}}$. For any $1\leq j\leq n$, and for
$k_{l+1}+1\leq s_{j}\leq k_{l}$ , such that $k_{a+1}=0$. Let
$s_{j}$ be the smallest number such that $[ \mathbf{B}_{s_{j},
j}] \neq 0$ and $\Lambda_{w_{j}}$ be the cross section of
$\Lambda$ in the coordinate system $W_{j} = \langle
e_{j}\rangle$. Then $\Lambda_{w_{j}}= 2{\Z}\ $  and  $\
P_{w_{j}}(\Lambda)={\Z}/2^{l-1}$.\\
\\
The proof is given in the appendix.\\
\\
\emph{Example 3.1:} Let $a=1 ,\alpha=2$ and $C_{0}, C_{1}$ are
two linear codes whose $C_{0}$ be the trivial code $\F^{7}_{2}$
and $C_{1}$ be the $(7,4)$ linear code``Hamming code'' thus the
lattice constructed using Construction $\textit{\textbf{D}}$ has
the following Generator and parity-check matrices:
\begin{equation}
\quad\ \mathbf{B}=
\begin{pmatrix}
\: 1 &\: 0 &\: 0 &\: 0 &\: 1 &\: 1 &\: 0 \: \\
\: 0 &\: 1 &\: 0 &\: 0 &\: 0 &\: 1 &\: 1 \: \\
\: 0 &\: 0 &\: 1 &\: 0 &\: 1 &\: 1 &\: 1 \: \\
\: 0 &\: 0 &\: 0 &\: 1 &\: 1 &\: 0 &\: 1 \: \\
\: 0 &\: 0 &\: 0 &\: 0 &\: 2 &\: 0 &\: 0 \: \\
\: 0 &\: 0 &\: 0 &\: 0 &\: 0 &\: 2 &\: 0 \: \\
\: 0 &\: 0 &\: 0 &\: 0 &\: 0 &\: 0 &\: 2 \: \\
\end{pmatrix},
\,\, \mathbf{B}^{*}=
\begin{pmatrix}
\: 1 &\: 0 &\: 0 &\: 0 &\frac{\:}{\:}\frac{1}{2} &\frac{\:}{\:}\frac{1}{2} &\: 0\: \\
\: 0 &\: 1 &\: 0 &\: 0 &\: 0 & \frac{\:}{\:}\frac{1}{2} &\frac{\:}{\:}\frac{1}{2}\: \\
\: 0 &\: 0 &\: 1 &\: 0 &\frac{\:}{\:}\frac{1}{2} &\frac{\:}{\:}\frac{1}{2} &\frac{\:}{\:}\frac{1}{2}\: \\
\: 0 &\: 0 &\: 0 &\: 1 &\frac{\:}{\:}\frac{1}{2} &\: 0 &\frac{\:}{\:}\frac{1}{2}\: \\
\: 0 &\: 0 &\: 0 &\: 0 &\: \frac{1}{2} &\: 0 &\: 0\: \\
\: 0 &\: 0 &\: 0 &\: 0 &\: 0 &\: \frac{1}{2} &\: 0\: \\
\: 0 &\: 0 &\: 0 &\: 0 &\: 0 &\: 0 &\: \frac{1}{2} \: \\
\end{pmatrix}
\nonumber
\end{equation}
\\
Corollary 3.1, implies that $s_{1}=1\ ,\ s_{2}=2,\ s_{3}=3,\ s_{4}=4,\ s_{5}=1,\ s_{6}=1,\ s_{7}=2\ $. Therefore $\Lambda_{w_{j}}= 2{\Z},\ P_{w_{j}}(\Lambda)={\Z}\ and\ |G_{j}|=2\ (j=1,\ldots,7)$.\\
\\
The following theorem, which generalizes Construction $\textit{\textbf{D}}$ to any collection of linear codes without condition $d^{(l)}_{min}\geq\frac{4^{l}}{\alpha}$, is proved. This theorem shows the relation between the performance of the lattice and the performance of it's linear codes.\\
\\
\emph{Theorem 3.2:} Let $C_{0}\supseteq
C_{1}\supseteq\ldots\supseteq C_{a}$ be a family of linear codes
with $C_{l}=[n,K_{l},d^{(l)}_{min}]$ and let
$\{\textbf{c}_{1},\ldots,\textbf{c}_{n}\}$ be linear independent
vectors in $\F^{n}_{2}$ such that
\[
C_{l}=\langle \textbf{c}_{1}, \ldots , \textbf{c}_{K_{l}}\rangle,\
l=1,\ldots,a.
\]
Also let $\Lambda$ be the corresponding lattice given by
Construction $D$. Then we have
\begin{equation}\label{eq:06}
\frac{1}{\alpha} min \{
d^{(1)}_{min},4^{-1}d^{(2)}_{min},\ldots,4^{1-a}d^{(a)}_{min},4
\}\leq d^{2}_{min}(\Lambda)
\end{equation}
\\
The proof is given in the appendix.\\
\\
\emph{Corollary 3.2:} The coding gain of the lattice constructed
using Construction $\textit{\textbf{D}}$ is
\begin{equation}\label{eq:07}
\frac{\frac{1}{\alpha}\ min\ \{
d^{(1)}_{min},4^{-1}d^{(2)}_{min},\ldots,4^{1-a}d^{(a)}_{min},4
\}}{4^{1-\frac{\sum_{l=1}^{a}K_{l}}{n}}}\leq \gamma(\Lambda).
\end{equation}
and if $d^{(l)}_{min}\geq 4^{l},\ l=1,\ldots,a,$\quad then
\begin{equation}\label{eq:08}
\alpha^{-1}4^{\frac{\sum_{l=1}^{a}K_{l}}{n}}=\gamma(\Lambda).
\end{equation}
\\
The proof is given in the appendix.\\
%%%%%%%%%%%%%%%%%%%%%%%%%%%%%%%%%%%%%%%%%%%%%%%%%%%%%%%%%%%%%%%%%%%%%%%%%%%%%%%%%%%
\section{SYSTEMATIC \textit{LDGM} LATTICES}
\label{Sec:CONCLUSION} In order to have a low iterative decoding
complexity, we need to have a low density Tanner graph
representation for the lattice \cite{banihashemi2001}. To achieve
this goal the new class of lattices ``\textit{systematic low
density generator matrix ($SLDGM$) lattices}'', from
systematically $LDGM$ codes is constructed. It is known that when
the girth (the length of the shortest cycle in the Tanner graph)
of code's increases, then the minimum distance of the code also
increases \cite{Tanner}. The progressive edge growth (PEG)
algorithm is an efficient method for constructing a Tanner graph
having a large girth by progressively establishing edges between
symbol and check nodes in an edge-by-edge manner \cite{PEG}. For
$SLDGM$ lattices Corollary 3.1 implies that $g_{i}=2^{l}$, for
some $l\in\{1,\ldots,a\}$. If $\alpha=1$ and $a=1$, then one-level
$SLDGM$ lattices are obtained, we denoted by $SLDGM^{1}_{n}$ and
if $a=2$ then two-level $SLDGM$ lattices derived
which are denoted by $SLDGM^{2}_{n}$.\\
The Generalized Min-Sum Algorithm For Lattices Constructed Using
Construction $\textit{\textbf{D}}$ is used to decode $SLDGM$
lattices {\cite{hassan_mehri}. The upper bound of decoding
complexity per iteration is:
\begin{equation}\label{eq:09}
N_{dec}\leqslant
n(g.d_{s}^{max}(d_{s}^{max}-1)+g^{d_{ch}^{max}}.d_{ch}^{max}(d_{ch}^{max}-1)+g-1)
\end{equation}
where $d_{s}^{max}$ and $d_{ch}^{max}$ are the maximum degree of
symbol-nodes and check-nodes in the Tanner graph of the lattice
respectively and $g_{i}\leq g$ $(i=1,\ldots,n)$. \\
\\
In the following tables the performance of $SLDGM_{n}$ lattices
compared with $LDPC$ lattices ($L_{n}$) \cite{8}. In these tables
$N_{D} = N_{dec}\times$\textit{(average number of iteration)} and
$M_I$ denote the decoding complexity and maximum number of
iteration, respectively. $P_{e}^{*}=(2/n)P_{e}$ denotes the
normalized error probability\cite{TAROKH}.
\begin{center}
Table.1\\
\begin{tabular}{c|c|c|c|c}
\noalign{\hrule height0.2pt}
$SNR_{db}$ & $N_{D}$ & $ M_I $& $P_{e}$&  $P_{e}^{*}$\\
\hline
$1$&$\leq3.87\times10^{5}$&$7$&$7.98\times10^{-1}$& $6.23\times10^{-3}$\\
$2$&$\leq3.03\times10^{5}$&$6$&$2.65\times10^{-1}$&$2.031\times10^{-3}$\\
$3$&$\leq2.70\times10^{5}$&$4$&$3.71\times10^{-2}$&$2.889\times10^{-4}$\\
$4$&$\leq2.45\times10^{5}$&$4$&$7.967\times10^{-3}$&$6.22\times10^{-5}$\\
$5$&$\leq2.14\times10^{5}$&$3$&$4.00\times10^{-4}$&$3.125\times10^{-6}$\\
$6$&$\leq1.5\times10^{5}$&$2$&$2.003\times10^{-5}$&$1.56\times10^{-7}$\\
\noalign{\hrule height0.2pt}
\end{tabular}
\\
performance of $SLDGM_{256}^{1}$
\end{center}
\begin{center}
Table.2\\
\begin{tabular}{c|c|c|c|c}
\noalign{\hrule height0.2pt}
$SNR_{db}$ & $N_{D}$ & $ M_I $& $P_{e}$&  $P_{e}^{*}$\\
\hline
$1$&$1.9\times10^{6}$&$35$&$5.807\times10^{-1}$& $4.536\times10^{-3}$ \\
$2$&$9.5\times10^{5}$&$36$&$1.209\times10^{-1}$& $9.453\times10^{-4}$ \\
$3$&$4.66\times10^{5}$&$35$&$1.759\times10^{-2}$& $1.375\times10^{-4}$ \\
$4$&$2.74\times10^{5}$&$35$&$2.396\times10^{-3}$& $1.872\times10^{-5}$ \\
$5$&$2.33\times10^{5}$&$32$&$2.46\times10^{-4}$& $1.92\times10^{-6}$ \\
$6$&$2.25\times10^{5}$&$28$&$1.20\times10^{-5}$& $9.375\times10^{-8}$ \\
\noalign{\hrule height0.2pt}
\end{tabular}
\\
performance of $L_{256}^{1}$
\end{center}
The performance of one-level type of these lattices show that for
decoding $SLDGM$ lattices we didn't need large number of
iteration as done as for $LDPC$ lattices. one-level $SLDGM$
lattices have almost the same performance like one-level $LDPC$
lattices. The performance of two-level type of these lattices at
the same dimension is proposed as follows:
\begin{center}
Table.3\\
\begin{tabular}{c|c|c|c|c}
\noalign{\hrule height0.2pt}
$SNR_{db}$ & $N_{D}$ & $ M_I $& $P_{e}$&  $P_{e}^{*}$\\
\hline
$1$&$\leq2.70\times10^{5}$&$13$&$7.10\times10^{-1}$& $5.54\times10^{-3}$ \\
$2$&$\leq2.38\times10^{5}$&$7$&$3.23\times10^{-1}$& $2.52\times10^{-3}$ \\
$3$&$\leq1.87\times10^{5}$&$6$&$8.02\times10^{-2}$& $6.95\times10^{-4}$ \\
$4$&$\leq1.46\times10^{5}$&$4$&$1.00\times10^{-2}$& $7.81\times10^{-5}$ \\
\noalign{\hrule height0.2pt}
\end{tabular}
\\
performance of $SLDGM_{256}^{2}$\\
\end{center}
\begin{center}
Table.4\\
\begin{tabular}{c|c|c|c|c}
\noalign{\hrule height0.2pt}
$SNR_{db}$ & $N_{D}$ & $ M_I $& $P_{e}$&  $P_{e}^{*}$\\
\hline
$1$&$2.76\times10^{6}$&$23$&$9.51\times10^{-1}$& $7.439\times10^{-3}$ \\
$2$&$2.11\times10^{6}$&$21$&$3.41\times10^{-1}$& $2.666\times10^{-3}$ \\
$3$&$1.20\times10^{6}$&$23$&$7.35\times10^{-3}$& $5.754\times10^{-5}$ \\
\noalign{\hrule height0.2pt}
\end{tabular}
\\
performance of $L_{256}^{2}$
\end{center}
The upper bound of decoding complexity for $SLDGM$ lattices is
lower than decoding complexity for $LDPC$ lattices. As mentioned
before maximum number of iteration for decoding $SLDGM$ lattices
is lower than it for $LDPC$ lattices. Two-level $SLDGM$ lattices
have relatively the same performance like two-level $LDPC$ lattices.\\
These results show that $SLDGM$ lattices have almost good
performance. The $LDPC$ lattices encoder has to calculate the
generator matrix. Not that unlike $\mathbf{B}^{*}$,
$\mathbf{B}=(\mathbf{B}^{*})^{-1}$ is not sparse matrix, in
general, so the calculation requires nonlinear computational
complexity. This not a desirable property because the decoder's
computational complexity is linear \cite{8}. A possible solution
is to produce $LDGM$ lattices which have linearly encoding and
decoding complexities.
%%%%%%%%%%%%%%%%%%%%%%%%%%%%%%%%%%%%%%%%%%%%%%%%%%%%%%%%%%%%%%%%%%%%%%%%%%%%%%%%%%%
\section{CONCLUSION}
\label{Sec:CONCLUSION} With a slight modification in the
structure of Construction $\textit{\textbf{D}}$, A new class of
lattices in terms of their generator matrix is proposed. Theorem
3.2 provides lower bound on minimum distance of the lattice in
terms of the minimum distance of its underlying codes. Corollary
3.2 shows the relation between the coding gain of the lattice and
its underlying codes parameters. It is shown that cross sections
and projections of this class of lattices can be derived
properly. In addition to low encoding and decoding complexities,
these class of lattices have an acceptable performance under
iterative decoding algorithm. The performance of the $LDGM$
lattices depends on the performance of their underlying $LDGM$
codes. It would be interesting to construct other class of $LDGM$
lattices. Such constructions would improve, or provide us with
different performance compared to the class of $LDGM$ lattices
presented here.
%%%%%%%%%%%%%%%%%%%%%%%%%%%%%%%%%%%%%%%%%%%%%%%%%%%%%%%%%%%%%%%%%%%%%%%%%%%%%%%%%%%
\section{APPENDIX}
\textit{Theorem 3.1:} The proof is given
in~\cite{sadeghi_refrence_5}.\\
\\
\textit{Corollary 3.1:} By the definition of cross section and
projection of a lattice, the result is a direct consequence of
\textit{Theorem 3.1}\\
\\
\textit{Theorem 3.2:} We have $\frac{1}{\alpha}\leq
d_{min}^{(0)}$, because $C_{0}=\F^{n}_{2}$. Consider
$\textbf{x}\neq 0 \in \Lambda$, without lost of generality choose
$k\geq 0$ such that $2^{k}\textbf{x}\in \Z^{n}$ and
$2^{k-1}\textbf{x}\notin \Z^{n}$. Let $k\leq a-1$, Eq(2) yields
there would be $l\in\{1,\ldots,a\}$ such that
$\alpha^{(l)}_{j}\neq 0 $ thus we could find $1\leq j\leq k_{l}$
such that $\textbf{c}_{j}\neq 0$ where
$\textbf{c}_{j}=(c_{j_{1}},\ldots,c_{j_{n}})$. Then there would be
$j_{1}\leq j_{m}\leq j_{n}$ which $c_{j_{m}}=1$. Since
$\textbf{c}_{j}\in C_{K_{l}}$ as a result of Euclidean norm
$\|\textbf{c}_{j}\|^{2}\geq \frac{d^{(l)}_{min}}{\alpha}$. It
follows that
\begin{center}
$(\frac{1}{2^{l-1}})^{2}\|\textbf{c}_{j}\|^{2}\geq
\frac{d^{(l)}_{min}}{\alpha}4^{1-l}$,
\end{center}
Hence
\begin{center}
$\|\textbf{x}\|^{2}\geq \frac{d^{(l)}_{min}}{\alpha}4^{1-l}$.
\end{center}
Let $k\geq a$ then $\|\textbf{x}\|^{2}\geq \frac{4}{\alpha}$.\\
\\
\textit{Corollary 3.2:} The proof is a direct consequence of the
\textit{Theorem 3.2} and the definition of coding gain of the
lattice.
%%%%%%%%%%%%%%%%%%%%%%%%%%%%%%%%%%%%%%%%%%%%%%%%%%%%%%%%%%%%%%%%%%%%%%%%%%%%%%%%%%%

\end{document}